\documentclass[conference]{IEEEtran}

\usepackage{cite}
\usepackage{amsmath,amssymb,amsfonts}
\usepackage{algorithmic}
\usepackage{graphicx}
\usepackage{textcomp}
\usepackage{xcolor}
\usepackage{url}
\usepackage{overpic}
\def\BibTeX{{\rm B\kern-.05em{\sc i\kern-.025em b}\kern-.08em
    T\kern-.1667em\lower.7ex\hbox{E}\kern-.125emX}}

\usepackage{hyperref}
\hypersetup{colorlinks=true,allcolors=blue,
pdfauthor = {Eric S. Enos and Daniel R. Herber},
pdftitle = {Using Hybrid System Dynamics and Discrete Event Simulations to Identify High Leverage Targets for Process Improvement in a Skill-based Organizational Structure},}

\usepackage[stretch=10]{microtype}

\begin{document}

\title{Using Hybrid System Dynamics and Discrete Event Simulations to Identify High Leverage Targets for Process Improvement in a Skill-based Organizational Structure}

\author{\IEEEauthorblockN{Eric S. Enos}
\IEEEauthorblockN{{Doctoral Candidate} \\ \textit{Department of Systems Engineering} \\
\textit{Colorado State University}\\
Fort Collins, CO, USA \\
\href{mailto:eric.enos@colostate.edu}{eric.enos@colostate.edu}}
\and
\IEEEauthorblockN{Daniel R. Herber}
\IEEEauthorblockN{{Assistant Professor} \\\textit{Department of Systems Engineering} \\
\textit{Colorado State University}\\
Fort Collins, CO, USA \\
\href{mailto:daniel.herber@colostate.edu}{daniel.herber@colostate.edu}}
}

\maketitle

\begin{abstract}

This paper is based on a case study of an IT organization in a large, US-based healthcare provider, and develops simluation models to identify areas for performance improvement. These organizations are often grouped into departments by technical skill and support both operational work (tickets) and project work (tasks) of various priorities. From a practical standpoint, resource managers and staff regularly manage all work as queued and assign / complete it based on the priorities of the day.
%
%
%
Using project and operational metrics from the case study organization, the hybrid model using both system dynamics and discrete event simulation developed through this research depicts the flow of work through a skill-based team as well as many of the key factors that influence that workflow, both positive and negative. Experience indicates that the interaction between project and operational work -- as well as between teams with differing skills -- entangles work queues and wait times within those queues in a way that rapidly scales in complexity as the number of interacting individuals and teams increases. Results from model simulation bear out this intuition. Scaling the models to accommodate multiple teams is a topic of future research.

\end{abstract}


\section{Introduction} 

Leaders of all types find it challenging to complete work in a timely and high-quality manner in a skills-based organizational model, although this has been treated most explicitly in the context of call centers \cite{stolletz_performance_2004,koole_routing_2003} it is common across IT and in other functions such as R\&D \cite{hematian_multiobjective_2020}. This paper is intended to help explain why and provide clues on how to better shepherd work to completion. In particular, senior leaders focused on demand and supply management can leverage the methods outlined here to understand the root causes of work delays and propose improvements to reduce them.

It is not unusual for a large IT organization to have many dozens of projects active at any given time, representing large numbers of active tasks to be completed within a schedule. Similarly, most organizations have a similar (or larger) number of active tickets representing incidents and requests for service. Each of these tasks and tickets can have a variable --- and sometimes changing --- priority for completion. The time frame of this analysis is deliberately assumed to be short to preclude the addition of staff. In this short time frame, leaders are limited to adjusting individual work priorities and shifting individual team members between different work items. See Mitchell \cite{mitchell_basic_1997} and the CPHIMS Review Guide \cite{society_himss_cphims_2021} for additional background.

Healthcare provider IT is commonly organized around specific technology skills, with teams supporting both project and operational work. As a result, individuals are routinely assigned to multiple project tasks as well as operational tickets. This intensifies the need for team members to stop and start work on any given task or ticket based on changing work priorities. Of course, individuals vary in skill level, and only the highest-skilled team members can complete every task or ticket that is assigned to the group quickly and with high quality. These complicate the queuing within the team as assignments are juggled between team members. Further, many tasks and tickets require multiple skills to complete (e.g., a network engineer, a server administrator, and a security analyst), which results in queuing of work as it passes between teams. All of this contributes to a high percentage of queue time relative to the work being performed, which is deadly to timeliness and customer satisfaction.

Discrete event simulation (DES) models the queuing environment with stochastic arrival and service times. DES enables predictive output based on multiple iterations of the model, and complex work routing decisions are enabled through entity attributes \cite{choi_modeling_2013}. Queues are explicitly used to manage operational work in IT, and implicitly used for project task as well \cite{jensen_dynamic_2006}. That said, DES by itself cannot express the full complexity of work management over time. System Dynamics (SD) is used to rigorously analyze the positive and negative influencing factors that impact the flow of work. Stocks in these models are directly analogous to DES queues. The SD model is built incrementally, adding influencing factors until a full picture of a single team is built and then extended to show the interactions between multiple teams with different skills. The differences between the methods is outlined by Morecroft and Robinson \cite{morecroft_explaining_2014}, but to fully characterize the organization in this research a combination of both approaches is used as justified by Brailsford \cite{brailsford_towards_2010}.

In this paper, the combination of DES and SD models are linked through the alignment of the entity attributes in the Mathworks SimEvents DES model with the influencing factors in the Vensim SD model. Changes to these attributes are iterated based on the output of each model. The ultimate goal is to identify the influencing factors with the highest contribution to process outcomes and possible actions to effect improvement.

The rest of the paper is organized as follows.
In Sec.~\ref{sec:theory}, the previous research on the use of SD and DES methods in modeling the organization's work management is explored. Section~\ref{sec:methodology} outlines the process for obtaining baseline data and building the models used in the research. Next, Sec.~\ref{sec:results} explores the behavior of the models and the insights taken from them. Finally, Sec.~\ref{sec:futureresearch} explores the applicability of the hybrid DES-SD modeling technique and areas of expansion and refinement.

\section{Theory}
\label{sec:theory}

A system is defined as ``a collection of elements and a collection of inter-relationships amongst the elements such that they can be viewed as a bounded whole relative to the elements around them'' \cite{cloutier_guide_2023}. Organizations are well researched as complex adaptive systems \cite{anderson_complexity_1999}, and exhibit varying levels of complexity through feedback loops, which are often poorly understood and can lead to highly non-intuitive outcomes during their operation \cite{forrester_principles_1990}. Systems thinking is a collection of methodologies that enable consideration of the ``whole'' system, including its constituent parts and interactions \cite{caulfield_case_2001}. Work management within the organization under study meets these criteria.

\subsection{Systems Thinking and System Dynamics}

System dynamics is a rigorous methodology that has been successfully used in various contexts for modeling the dynamic behavior of complex managerial and organizational systems such as those considered here \cite{jonkers_connecting_2021, abdel-hamid_dynamics_1984, groundstroem_using_2021, homer_system_2006}. The models referenced in the sections below each address separate aspects of the work dynamic within a healthcare IT organizational model, but none fully explore all types of work commonly serviced by these teams and only partially identify the leverage points which improvements can potentially address. As such, they will be used as the basis for a consolidated model that better highlights these areas. These are each expanded below, along with their contribution to the model for this study.

\subsection{Dynamics of Project Management}

Scheduled work in the form of projects has received substantial attention from researchers with an interest in systems dynamics. Projects -- especially large projects -- have been long recognized as highly complex internally, as well as in relation to the rest of the organization \cite{san_cristobal_complexity_2018, baccarini_concept_1996}. There is a rich body of research on the applicability of system dynamics as applied to project management of single projects \cite{lyneis_strategic_2001, rodrigues_system_1997, rodrigues_role_1996, abdel-hamid_lessons_1989}. Lyneis and Ford, in particular, developed this model depicting the management of scheduled work through several iterations\cite{lyneis_system_2007, ford_project_2007}. Ordonez et al.\cite{ordonez_study_2019} elaborated on the characteristics of a multi-project environment that apply to project managers, functional managers, and staff, including the need for staff to multitask between projects. Platje and Seidel\cite{platje_breakthrough_1993} emphasize the complexity of balancing costs, resource allocations, and completion times in these scenarios, while Van Der Merwe\cite{van_der_merwe_multi-project_1997} explores the interplay between functional and project managers in managing work. Payne estimates that up to 90\% of all projects are run in this context, and often lead to complex, matrixed organizational structures\cite{payne_management_1995}. These characteristics are observed in the case study organization in Sec~\ref{sec:methodology}.

Kang and Hong\cite{kang_evaluation_2009} explain the competition for limited resources between projects and the resulting increases in queue time as each project waits for resource availability, even with close attention to resource allocation.  This dynamic highlights the importance of reducing queue time to accelerate project delivery. Patanakul and Milosevic\cite{patanakul_competency_2008} discuss the unique demands on project managers who manage multiple efforts concurrently, which often have unrelated goals and stakeholder needs. Diao and Hecheng\cite{diao_modeling_2011} acknowledge similar management overhead in the context of coordinating operational tickets between teams. These issues certainly apply to the functional leaders in the organization under study as well as project managers.

Platje and Seidel\cite{platje_breakthrough_1993} discuss the need for operational managers to delegate more to subordinates in conditions of high operational uncertainty, such as that created by the need to support multiple work types and priorities. Rahmandad and Weiss\cite{rahmandad_dynamics_2009} emphasize the interactions between projects and the need to develop ``slack'' in resource capability to be able to absorb changes in priorities and demand, and warn that tipping points exist with sustained schedule pressure.

Finally, Jensen et al.\cite{jensen_dynamic_2006} developed a model depicting the interactions between ``work stacks'' -- which could be between individuals focused on incidents (repair/reactive work) and on project tasks (maintenance/proactive work), between differently skilled teams, or both. This bridges the gap between project and operational work outlined in the next section.

\subsection{Dynamics of IT Service Management}

The prevalent framework for managing work that arises based on events that occur within the organization, generally categorized as ``incidents'' and ``requests'', is the IT Information Library (ITIL). Incident and request management in IT organizations is routinely managed through queue-based ticketing systems, with queues assigned to individuals and teams. These were successfully modeled as queuing systems by Bartolini et al. using their SYMIAN simulation \cite{bartolini_symian_2010} and treated by other researchers\cite{antoniol_queue_2001}. ITIL was developed in Great Britain in the 1980's and has served as the de-facto standard for IT infrastructure and operations since the publication of version 2.0 in 2000. Version 3.0 was released in 2007, and version 4.0 was released in 2019. All versions of ITIL since 2.0 have treated the management of incidents and requests as a queueing problem. 

Voyer et al. \cite{voyer_system_2015} developed a model of major incident management that can be used as a basis for one major type of unscheduled work. Wiik and Kossakowski \cite{wiik_dynamics_2017} developed a model of incident management that specifically incorporates the benefits of automation as applied to information security response activities. Neither of the models above reflects differentiation of skills, either in terms of skill \textit{level} or skill \textit{type}, and therefore do not address the routing of tickets between individuals or teams due to incorrect initial assignment or the need for multiple teams to collaborate to complete the ticket. Discussion of the dynamics of a multi-level (skill) service desk operation is discussed in Fenner et al. \cite{fenner_system_2015}, and treatment of these issues resulting in ticket re-routing / reassignment is addressed by Li et al. \cite{li_measuring_2013}.

Oliva developed a model of request management that can be used as a basis for the second major type of unscheduled work \cite{oliva_dynamic_1996}. While not explicitly treated in their model, note that the concept of ``work pressure'' as represented can be interpreted as impacting the relative \textit{priority} of work items. This is an important consideration in the assignment of tickets in any operational model as discussed by Li et al. \cite{li_it_2010} and can be fruitfully extended to project tasks as well.

\subsection{Discrete Event Simulation}

Discrete event simulation (or DES) is a fundamentally different approach to process modeling based on queuing theory. models essentially depend on several key concepts: \textit{entities} (along with attributes that can be assigned to entities), \textit{resources} (such as queues and servers that act on entities), and \textit{activities} (including routing between resources based on attributes and action taken by servers). Further, \textit{attributes} track any changes in entity state that occur during specific \textit{events} (such as entry into a queue or completion of service) \cite{choi_modeling_2013}.

A key distinction between DES and system dynamics is in the word ``discrete'': each entity is distinct, and events happen at discrete points in time, while system SD continuous flow through the model controlled by rates of change, which are determined by differential equations \cite{brailsford_comparison_2001}. Another element that sets DES apart from SD is that it is inherently stochastic in the assignment of key elements of the model \cite{chahal_conceptual_2013}. Key model settings such as inter-arrival times, service times, and if needed the values of entity attributes are determined through probability distributions. Finally, DES models are considered to be predictive, with complex queuing and routing systems displaying emergent behavior over many iterations, while SD models are considered descriptive of the effect of causal loops on the underlying queuing system.

The applicability of DES to operational work management is obvious, as operational tickets are commonly managed through explicit queuing systems -- and from a practical standpoint, project tasks can also be considered to be queued, as discussed below. DES allows the construction of highly valid and verifiable models of the management of work in environments such as the case study organization. 

\subsection{Hybrid Modeling}

These models can be used together to retain the accuracy and predictive capabilities of statistical queuing within DES models while also adding the broader descriptive capability of the SD model \cite{gonul-sezer_comparison_2016} \cite{chahal_conceptual_2013}. For the purposes of this analysis, the goal is to obtain a clear understanding of the effects of the dynamical influences on the queue time and total throughput time of entities in the DES model \cite{greasley_simulating_2019}.

While it is theoretically possible to couple these models programmatically (either by integrating two separate tools or by leveraging a tool such as AnyLogic, which inherently enables both model types) in this research, MathWorks SimEvents is used for DES while Vensim is used for SD modeling, and data is passed between them manually following each tool's simulation run. This is referred to as \textit{cyclic interaction} in \cite{chahal_conceptual_2013} as opposed to \textit{parallel interaction}.

Hybrid modeling requires the explicit modeling of key influences in the SD model as entity attributes in the DES model, so that these can be adjusted in subsequent iterations of the models based on the results of previous simulations.


\section{Methodology}
\label{sec:methodology}

\subsection{The Case Study Organization}

The case study organization is a nationwide healthcare provider in the United States with a large number of acute- and non-acute care hospitals, ambulatory clinics, and physician practices. The application environment is a combination of on-premises and Software-as-a-Service (SaaS)-based systems with a substantial physical infrastructure. At the time of this study, central IT consisted of roughly 700 staff, organized by technical skill (network, security, etc.) and function (project management, application analyst, etc.). Operational requests and incidents are managed in ServiceNow (a leading ticketing system), and projects are managed through several different applications and tools. There were several hundred active projects of various sizes underway, with dozens of tickets of both types varying complexity arriving daily.

Large healthcare providers in the United States are generally assumed to share the following characteristics that affect the structure of their IT organizations:

\begin{itemize}
\item Relatively low margins with increasing erosion due to evolving industry dynamics driving high rates of mergers, acquisitions, and divestitures and resulting high levels of system variability and high pressure to reduce staff costs.
\item Earnings-driven incentive systems that discourage operational costs in favor of capital investments and drive:
    \begin{itemize}
    \item Significant dependence on capitalizable, on-premises COTS applications and commensurately on long-lived, ``mutable'' systems.
    \item Commensurately low reliance on cloud-based infrastructure, systems, and associated work management methods (e.g., DevOps).
    \item Prevalence of waterfall project methods over agile approaches (due to the above factors).
    \end{itemize}
\item High and increasing operational and clinical dependence on IT system and data performance and availability as direct contributors to clinical quality and safety, which has driven:
    \begin{itemize}
    \item Ubiquitous adoption of ITIL 2/3, especially for help desk, incident, and request management processes.
    \item Increasing number and complexity of projects focused on maintaining regulatory compliance, gaining cost efficiencies, improving the quality of care, and supporting innovation.
    \end{itemize}
\end{itemize}

These combine to create a complex technical environment and organizational structure, with a large volume of operational tickets and the need to simultaneously pursue a variety of projects ranging from routine to critical transformational efforts. The conclusions drawn from this research may have broader applicability to organizations or industry segments with similar characteristics.

The models developed in this paper only depict interactions within a single team. The modeled team supports mixed work types (e.g., both unscheduled and scheduled). Individuals can be primarily assigned to one or the other work type but can work on either if priorities necessitate at any given time. While researchers have demonstrated the preference to protect scheduled work from unscheduled demands, this is not always economically feasible. Each team supports multiple concurrent projects / products at different stages of planning and execution, as well as multiple concurrent incidents and requests. Due to the limited number of resources with particular skills, work of both types is queued awaiting completion. The models are intended to reflect the flow of work over "relatively" short timelines and does not address the ability to flex staff through hiring (or contract outsourcing) within the time window under analysis. The period of validity for the models is assumed to be roughly six months.

Resources may have specific tasks and/or tickets assigned to them in some cases and may, in other cases, pull work from a team queue based on perceived priorities. The assigned resource or a manager can make the decision to stop or reassign any piece of work for the reasons elaborated below. Active work in progress can be returned to the queue due to 1) requiring a higher skill level than the assigned resource; 2) being interrupted by higher priority/urgency work; or 3) being ``reassigned'' to another team's queue due to a lack of certain technical skill in the originating team. Note that this can happen multiple times with a given piece of work, even within a single team; when multiple teams are involved, a work item can spend significantly more time in queue than being actively worked. 

\subsection{The System Dynamics Model}

The model summarized in this work is built using Vensim PLE. It is intended to incorporate key elements of previous models that specifically highlight the areas where automation may be of benefit. As discussed above, the time horizon for the analysis is too short to allow for adjustments to resource availability through new hires or sourcing arrangements. 

Beginning with project work, the base loops and flows are shown in green, where the boxes represent the primary flow of project work, the green flows represent the ``happy path'' of work through the system, and the red flow represents work that is returned to the queue (work stoppages in this diagram). Regarding arrows, red arrows represent negative influences on the throughput of work through the process, green arrows represent positive influences, and blue arrows are neutral. The light green boxes indicate key measures, while the yellow boxes indicate where deviations of actual measures (vs. targets) generate managerial actions to improve productivity or quality.

\subsection{Full Single-Team Model}
\label{sec:fullsingleteam}

\begin{figure*}[t]
\centering
\includegraphics[width=0.9\linewidth]{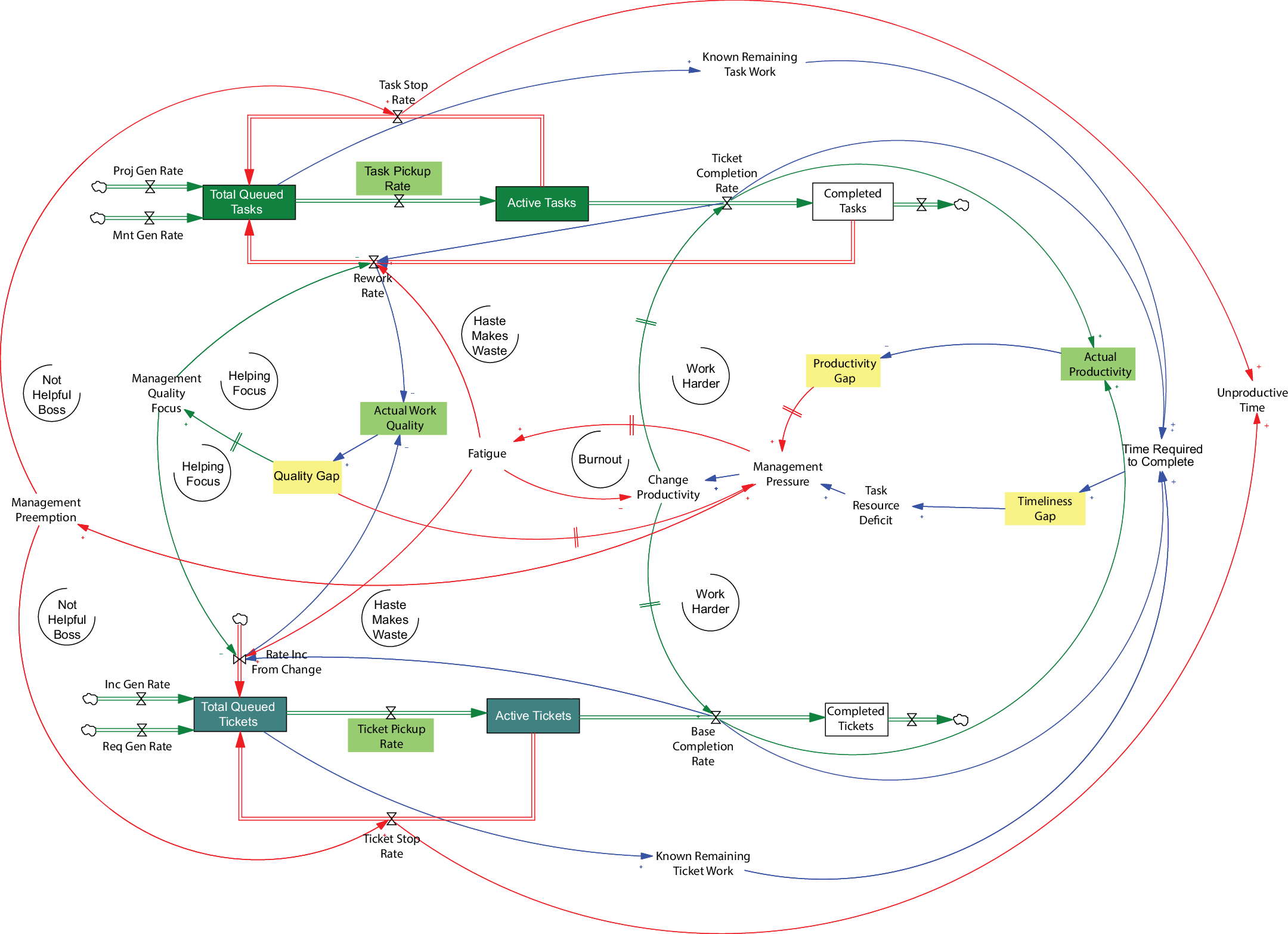}
\caption{Vensim system dynamics model of a single team reflecting both project and operational workflows.}
\label{fig:4switchcostprodv2}
\end{figure*}

Subsequent elaborations add the following aspects until reaching the full single-team model in Figure~\ref{fig:4switchcostprodv2}:
\begin{itemize}
\item Both operational work and rework (including incidents from change), including interactions between the different work types. The operational work is shown below the project work in teal boxes.
\item The effects of management response, the introduction of the possibility of a gap between desired and actual work quality, with a delayed response by management that can either be positive (through constructive assistance) or negative (through pressure causing fatigue).
\item The effects of resources having to stop work on a particular task / ticket and shift to another introduces the concept of switching costs which have a negative impact on overall productivity. 
\item The influence of ``timeliness'' corresponding to the on-time delivery of project tasks and the rapid fulfillment of operational tickets.
\end{itemize}

\subsection{The Discrete Events Simulation Model}

By comparison to the system dynamics model, the DES model is superficially much simpler; however, the complexity is embedded in the entity attributes and routing rules based on them.

\begin{figure*}[t]
\centering
\includegraphics[width=\textwidth]{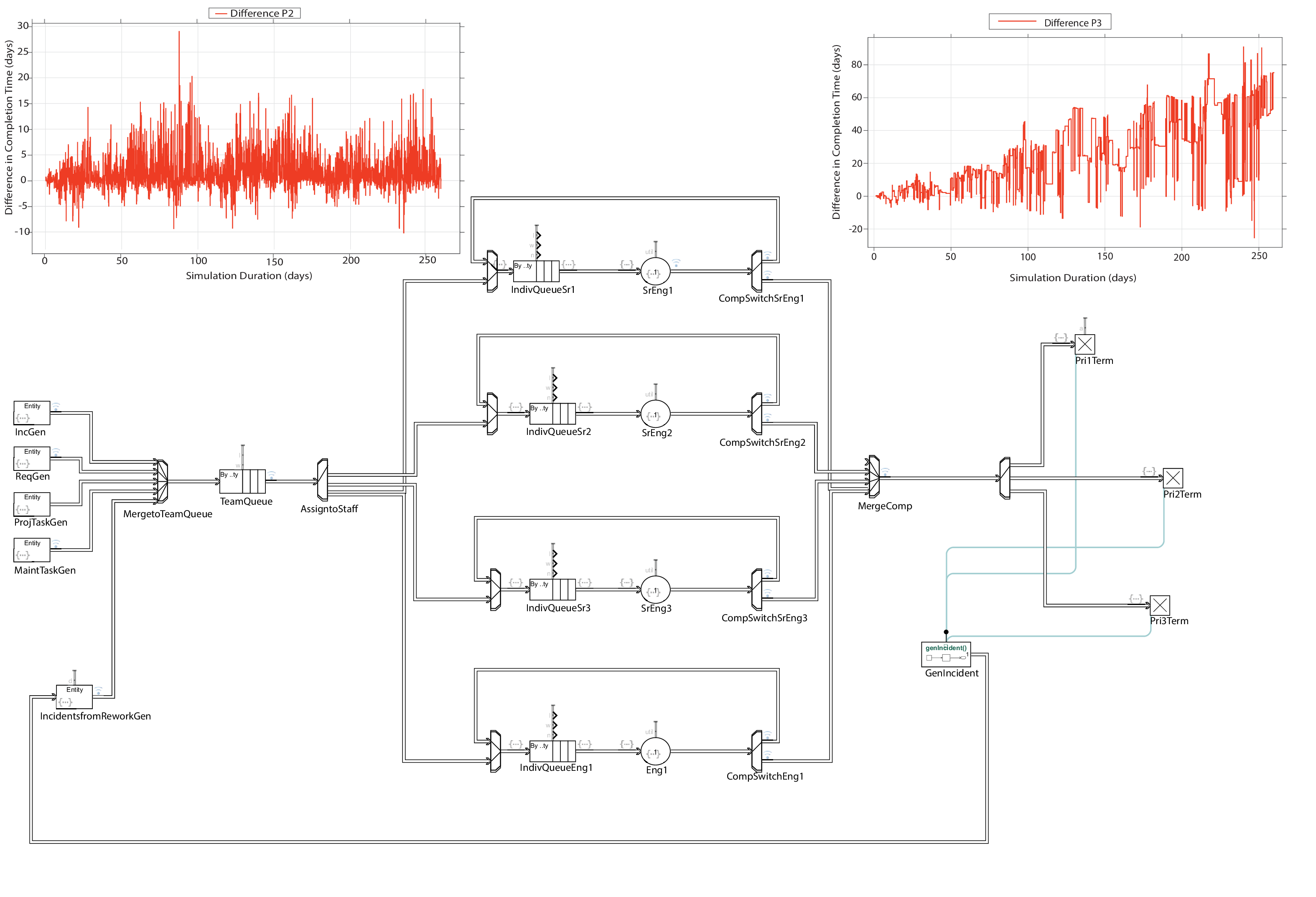}
\caption{Simevents queuing model depicting entity generators for each work type feeding four engineers with individual queues.}
\label{fig:DESmodel}
\end{figure*}

Greasley recommends clearly defining the scope of a DES model, including assumptions, abstractions, and areas deliberately left out of scope \cite{greasley_simulating_2019}. In order to maintain a manageable level of complexity, no additional teams are included; however, this is an abstraction as, in reality, several teams can be involved in even relatively simple and frequent tickets or tasks. The teams are modeled with accurate staffing in terms of numbers, skill level, and skill type of team members. These team members are modeled as \textit{servers} in SimEvents. Note also that this model is the more appropriate place to deal with the issues of (re-)prioritization and the impact of skill level mismatches between the task / ticket and the assigned resource on error rates, as these issues are more complicated to model in Vensim. The model is shown in Figure~\ref{fig:DESmodel}, with work generators on the left, a team queue to consolidate the work types, and four parallel individual queues and engineer ``servers''. Work stoppages are sent back to the individual queues, and completed work is forwarded to the termination points on the right. Note that the probability of error generation is captured through a signal from each termination point and generates incidents that are put back into the team queue.

The data to support determination of the inter-arrival and service time data for operational tickets is derived from six months of actual service desk system data and then fit to specific Poisson distributions using Palisade Software's @Risk for each team and ticket type. Data related to tasks was estimated through interviews with department leaders and also modeled as Poisson distributions. Tickets and tasks of different types are modeled as entities in SimEvents, with independent generators driven by appropriate distribution settings. Baseline data related to distributions of skill type and level required to complete tickets / tasks and other attributes that drive statistics of routing are based on estimates from interviews with department leaders. The different work generators create events (tasks) with the appropriate distribution of priority (P1 = high, P2 = medium, and P3 = low) to match the fitted data for that team and work type.

\subsection{Interaction Points Between Models}

As recommended by \cite{greasley_simulating_2019}, the following interaction points are defined between the DES and system dynamics models:

\begin{itemize}
\item Completion, rework, and preemption rates from the SimEvents model are fed forward into the Vensim model by adjusting the associated work generation rates in the stock-and-flow diagrams.
\item The effect of fatigue driven by increasing work intensity in the Vensim model results in an increasing probability of rework and incidents from changes over the simulation time, which is fed back into the SimEvents model directly.
\item The effect of increasing management pressure in the Vensim model is fed back into the SimEvents model as an increasingly frequent interruption of in-process work (i.e., having to stop a task / ticket), modeled through a proportional decrease in the engineers' available service time.
\end{itemize}

The cycle is iterated to determine the changes in the performance of the queuing process under the influence of changing dynamics. These changes are finally analyzed to determine the impact on the model (in terms of the completion rates) on the dynamic attributes driven by management interventions and resource responses to changes in pressure over time, as well as the sensitivity of the changes to changes in specific attributes.

\section{Results and Discussion}
\label{sec:results}

The base SimEvents model demonstrates that the team can adequately and quickly handle high- and medium-priority tasks and tickets (within a day and two working weeks, respectively), but that low-priority work completion times continue to increase these queues build monotonically throughout the simulation (ranging from 140--220 days by the end of the simulation, with an average of 88 days). This is consistent with observations from historical ServiceNow data -- the department analyzed is not in equilibrium at the start of the simulation.

While the Vensim model does not represent the differences in priority, the same steady increase in queuing is observed. The additional influences introduce different behaviors over time, including oscillations in quality, timeliness, and productivity that flow through to the observed behavior of the queues and flow rates. 

The dynamic behavior observed in the error generation and stop rates has a strong impact on subsequent iterations of the SimEvents model. After feeding back the changes from the Vensim results to the SimEvents model in the second iteration: 

\begin{itemize}
\item 1/3 more work items were stopped and re-queued due to management pressure, and there was a very large increase in Incidents from Rework. 
\item This resulted in a 25\% increase in work completed to reactive incident response (because there were so many more of them) as well as an increase in all completion times by 18\% for P1s and a 125\% increase in P2s.
\item For all intents and purposes, many P3s simply remained queued with 75\% less completed during the simulation. 
\item These differences are shown over time in the Difference graphs in the top corners of Figure \Ref{fig:DESmodel}) -- the graphs show the increase in days to complete work over time between the baseline and the next iteration. For medium-priority (P2) work items in the top left graph, the majority of data points are clustered above the 0 line, indicating longer completion times between simulation iterations. For low-priority  work items in the top right the completion times are not only longer but steadily increasing throughout the second run.
\end{itemize}

In essence, these interactions create a new reinforcing loop between the model iterations that drive increasing queue times, especially for medium- and even high-priority work.

\subsubsection{Improvement Focus Areas}
\label{sec:improvetgts}

The models reinforce several common-sense targets for improvement, such as reducing \textit{completion times} and improving \textit{responsiveness} for operational tasks, improving actual \textit{work quality} and increasing \textit{pickup} and \textit{completion rates}. More surprisingly, the hybrid model predicts that reducing the rate of \textit{rework} and the generation of \textit{new incidents} and reducing distractions (including preemption caused by managerial pressure) that interrupt work completion and increase \textit{switching costs} will have strong effects on work throughput.

\subsubsection{Strategies for Improvement}
\label{sec:improvstrat}

The models indicate that the interaction between project and operational work -- and likely between teams with differing skills -- creates a coupling of work queues and wait times within those queues: each shift of tasks / tickets within and between queues adds queue time to the work. The following strategies can be used independently and in combination to improve the outcomes with respect to the improvement targets outlined above:

\begin{itemize}
\item Ensure close coordination between project and functional leaders. 

\item Limit the number of projects in process through a governance and prioritization function -- i.e., portfolio management. 

\item Separate operational and project responsibilities between different staff within each department to reduce the amount of work transferred between those work type queues. 



\item Automate high-volume and repeatable work in order to improve response and task completion times and allow staff to focus on unique activities. Automation can also improve quality (assuming adequate testing) by ensuring consistently accurate outcomes, which is especially important for tasks that happen regularly or at scale. 

\end{itemize}

The increased use of automation, in particular, can address several improvement areas simultaneously, as demonstrated by its use in organizations leveraging DevOps-style methods - particularly the set of practices known as Continuous Integration / Continuous Deployment (CI / CD). 

\subsubsection{Key Metrics}
\label{sec:key metrics}

Finally, the model identifies several key metrics for leaders to understand delivery performance and which of the strategies above may provide the biggest improvements in performance at any given point in the organization's journey. Again, several of these are reasonably straightforward, such as the difference between actual and expected service delivery quality, in terms of both fitness for purpose and fitness for use; the difference between actual and desired staff / team productivity, as measured by task and ticket completion rates as well as the amount of rework created. 

Other metrics are less obvious but perhaps more easily managed and include: reassignment and re-queuing counts used as indicators of how many times work has been started and stopped; the number of tickets / tasks assigned to teams and individual staff as a proxy for understanding the amount of ``work juggling'' and resulting switching costs; and finally the rate of errors resulting in rework and or new incidents resulting from the previous changes.

\section{Conclusions and Future Research}
\label{sec:futureresearch}

The hybrid use of systems dynamics and discrete event simulation allowed a deeper exploration of the characteristics of priority queuing and management pressure by leveraging the strengths of each tool -- more so than would have been possible in either tool exclusively. Further, the process of defining which data is passed between tools -- and in what manner -- sharpens the understanding of what attributes are truly important in the underlying system. The use of coupled models exposes that increasing levels of rework and re-prioritization (expediting work) generate a reinforcing loop that degrades performance on low-priority work over time in the absence of mitigating strategies. This would not have been visible using either model alone.

Future research will add the unique interactions between two skill-based teams, as the inter-team interactions become much more complex as additional teams become part of the work process. This would be the case with a real cross-functional process such as server provisioning, which can cross multiple departments, technologies, and tools. The modeling of two teams is anticipated to provide a flavor of these interactions, which scale with the number of team pairings as $n(n-1)/2$. Modeling these interactions will allow us to address the potential impact of hiring and (and training) multi-skilled resources and creating cross-functional teams - both of which can prevent transfers of work between department queues.

Ultimately, this research can be leveraged to justify the expanded use of automation within IT organizations and assist leaders in the identification of use cases for automation that will have the greatest impact to performance.

\section*{Acknowledgments}
Special thanks are due to Steven Conrad and Kamran Eftekhari-Shahroudi at Colorado State University for their assistance with the model-building process.

\bibliographystyle{myIEEEtran}
\bibliography{references_syscon}

\vspace{12pt}

\end{document}